# Computational Insights into the Chemical Reaction Networks of $C_3H_6O_3$, $C_3H_7O_3$ and $C_2H_5O_2$: Implications for the Interstellar Medium

Anxo Lema-Saavedra,[a] Antonio Fernández-Ramos,[a,b] and Emilio Martínez-Núñez*[b]

**Abstract**

The formation of complex organic molecules (COMs) in the interstellar medium (ISM) is central to astrochemistry and prebiotic chemistry, as these species may act as precursors to biomolecules essential for life. Among COMs, glyceraldehyde ($HOCH_2CH(OH)C(O)H$, GCA) has attracted attention as a potential building block in early biochemical pathways. Although GCA has not yet been detected in the ISM, the presence of structurally related compounds in various astronomical environments suggests that it may form under interstellar conditions. In this study, we employed the automated reaction discovery tool AutoMeKin to systematically explore the gas-phase chemical reaction networks (CRNs) of $C_3H_6O_3$ (GCA), $C_3H_7O_3$ (a hydrogenated analog), and $C_2H_5O_2$. Reaction pathways were characterized at the ωB97XD/Def2-TZVPP level of theory, and rate coefficients for key processes were computed using the competitive canonical unified statistical (CCUS) model, which accounts for multiple dynamic bottlenecks. Our analysis revealed several barrierless pathways leading to GCA or to GCA and a leaving group. Notably, the reaction between glyoxal (HCOHCO) and the $HOCHCH_2OH$ radical, though neither has yet been detected in the ISM, was found to efficiently produce GCA and a formyl radical, with rate coefficients on the order of $5.4 - 7.9 \times 10^{-10}$ cm³ molecule⁻¹ s⁻¹ across the 10–100 K temperature range. However, aside from the aforementioned exception, most GCA formation channels result in highly vibrationally excited intermediates that are more likely to undergo rapid unimolecular decomposition than to be stabilized by radiative emission under typical ISM conditions. These results suggest that while gas-phase GCA formation is chemically feasible, it is likely transient and difficult to detect directly. In contrast, alternative products such as formaldehyde, glycolaldehyde, and (Z)-ethene-1,2-diol dominate many pathways and align better with current astronomical observations. This work provides detailed mechanistic and kinetic insights that enhance astrochemical modeling and advance our understanding of molecular complexity in star-forming environments. Furthermore, it highlights the utility of automated CRN exploration for uncovering viable synthetic routes to prebiotic molecules in space.

## 1 Introduction

The formation of complex organic molecules (COMs) in the interstellar medium (ISM) is a key area of research in astrochemistry, as these molecules are considered precursors to life. Among COMs, glyceraldehyde ($HOCH_2CH(OH)C(O)H$, GCA), the simplest sugar, has garnered attention due to its potential role in prebiotic chemistry and its relevance as a precursor to biologically relevant molecules.[1,2]

The gas-phase decomposition of GCA has been investigated experimentally through pyrolysis,[2] and the early steps of the Formose mechanism have been studied computationally using density functional theory (DFT).[1] Under pyrolytic conditions, GCA starts decomposing at 400°C, undergoing retro-aldol fragmentation and dehydration. The retro-aldol reaction predominantly yields glycolaldehyde and formaldehyde.[3,4]

Although GCA has not yet been detected in the ISM despite dedicated observational searches,[5] its structural analogs, such as glycolaldehyde and ethylene glycol, have been observed in star-forming regions and comets,[6-9] suggesting that it may also form under astrochemical conditions. Laboratory experiments and computational studies indicate that GCA may originate from formaldehyde-based reactions in interstellar ice analogues, driven by ultraviolet irradiation, radical recombination, or thermal processes,[2] or from the reaction of (Z)-ethene-1,2-ediol with hydroxycarbene (HOCH).[10] Understanding the formation pathways of GCA and related species in the ISM provides valuable insights into the chemical processes that drive molecular complexity and, ultimately, the emergence of life's building blocks in space.

This study explores potential gas-phase formation pathways for GCA, as well as other key reactions relevant to astrophysical environments, evaluating their viability under interstellar conditions. The formation of COMs in the ISM can proceed through various reaction mechanisms. Given the extremely low temperatures of interstellar clouds (10–150 K), gas-phase reactions must generally proceed without significant activation barriers.[11] In other words, such reactions are expected to follow energetically favorable pathways with submerged barriers,[12,13] often involving at least one radical or ion species. Additionally, due to the extremely low densities of interstellar clouds, product stabilization is inefficient unless radiative cooling provides a competitive mechanism.[11]

In particular, we investigate two types of GCA gas-phase formation reactions:

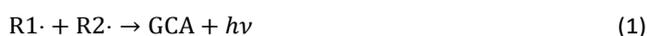

$$R1\cdot + R2\cdot \rightarrow GCA + h\nu \qquad (1)$$

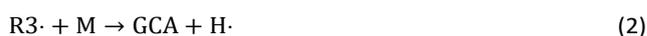

$$R3\cdot + M \rightarrow GCA + H\cdot \qquad (2)$$

Reaction (1), known as radiative association, involves two radicals (R1· and R2·) and requires efficient radiative stabilization of GCA. In contrast, reaction (2) is a radical-molecule reaction that does not rely on radiative stabilization, as it produces GCA and atomic hydrogen as separate products.[11]

Additionally, we also examined the possibility of reactions analogous to (2), where the leaving group is a small radical rather than H·.

To explore these reaction mechanisms, we employ AutoMeKin,[14-16] an automated tool for discovering chemical reaction networks (CRNs). AutoMeKin has been successfully used to propose gas-phase formation routes of various interstellar molecules, including benzene,[17] cyanoketene,[18] cyanamide,[19] formaldehyde,[20] glycolonitrile,[21] hydrogen isocyanide,[22] indole,[13] malononitrile,[23] vinyl alcohol,[24] and (Z)-1,2-ethenediol.[25]

In this study, we employ AutoMeKin to construct detailed reaction pathways for species with the chemical formulas $C_3H_6O_3$ and $C_3H_7O_3$, corresponding to reactions (1) and (2), respectively. $C_3H_6O_3$ represents GCA, while $C_3H_7O_3$ corresponds to GCA with an additional hydrogen (GCA-H). Additionally, we investigate the $C_2H_5O_2$ CRN using AutoMeKin due to its close relationship with the other CRNs. Through the analysis of these networks, we identify key intermediates and reaction channels, and we perform kinetic calculations to determine rate coefficients for the most relevant reactions. The computed rates provide valuable inputs for astrochemical models.

Our findings have the potential to bridge gaps in current astrochemical models by providing accurate rate coefficients and detailed reaction mechanisms. As observational techniques continue to advance, the insights gained from this work will contribute to a deeper understanding of the chemical evolution of star-forming regions and the origins of molecular complexity in space.

## 2 Computational details

### 2.1 Automated mapping of reaction mechanisms

The CRNs of $C_3H_6O_3$, $C_3H_7O_3$ and $C_2H_5O_2$ were generated using AutoMeKin,[14-16] an automated software developed in our group. This open-source program enables an efficient exploration of molecular decomposition pathways by integrating reactive MD simulations, graph-theoretic methods, and interactive visualization tools. The workflow begins with a preliminary potential energy surface (PES) exploration using semi-empirical calculations (hereafter referred to as Level1), followed by refinement of stationary points with a higher-level DFT or ab initio method (hereafter Level2). Consistent with previous studies, Level1 corresponds to PM7,[26] while Level2 employs ωB97XD/Def2-TZVPP. The latter has been demonstrated to accurately estimate barrier heights,[27] with Gaussian09 utilized for these calculations.[28] Additionally, AutoMeKin now incorporates an improved algorithm for identifying barrierless reaction pathways.[13]

Beyond its core capabilities in MD simulations and graph-theoretic analysis, AutoMeKin has recently been extended with a Python-based library, *amk_tools*, designed for parsing, processing, and analyzing CRNs.[13] Custom Python scripts further allow for the adaptation of energy profiles generated by *amk_tools* and facilitate the comparison of the predicted decomposition products with known interstellar molecules.

For the $C_3H_6O_3$ and $C_3H_7O_3$ molecular systems, the AutoMeKin workflow was executed iteratively 30 times, with each cycle comprising MD simulations, optimization of minima and transition states (TSs), and CRN generation. Each MD simulation set included 500 trajectories, with a maximum simulation time of 0.5 ps. For the smaller $C_2H_5O_2$ system, the workflow was run for 20 iterations with 480 trajectories per cycle.

The transition-state filtering criteria from our previous study on indole decomposition[13] were applied to remove structures with very low imaginary frequencies and eliminate redundant TSs. Additional details on the methodological approach are provided in Ref. 13.

To evaluate the convergence of the CRNs obtained through iterative refinement, kinetic Monte Carlo (KMC) simulations were conducted to solve the chemical master equation.[29] Specifically, the abundances of decomposition products were determined by exciting $10^3$ molecules of GCA and GCA-H at 250 kcal/mol. The reaction rate coefficients were estimated using RRKM theory,[30] given by:

$$k(E) = \sigma \frac{W^\ddagger(E-E_0)}{h\rho(E)} \qquad (3)$$

where, $\sigma$ represents the reaction coordinate degeneracy, $W^\ddagger(E - E_0)$ is the sum of states of the TS, $E_0$ is the barrier height, and $\rho(E)$ is the density of states of the reactant. These calculations were performed using AutoMeKin's kinetics module, applying the Beyer-Swinehart algorithm[31] for direct counting of sums and densities of states. The product abundances obtained across iterative steps are shown in Fig. S1, indicating convergence after 10-15 iterations. Notably, water, glycolaldehyde, formaldehyde, and their isomers are among the predominant decomposition products of GCA, consistent with previous pyrolysis studies.[3, 4]

While our study primarily focuses on the barrierless formation pathways of GCA, the complete CRNs are available in Zenodo.[32] In these networks, minima, transition states, and products are labeled as MIN, TS, and PROD (or PR), respectively, with numerical identifiers distinguishing different species. The labels for each network are assigned independently. Molecular visualizations and representations were performed using CYLview.[33]

### 2.2 Kinetic simulations

As in our previous study on glycolonitrile,[21] we apply the competitive canonical unified statistical (CCUS)[34, 35] model to reactions that exhibit two dynamic bottlenecks. Specifically, the first bottleneck usually corresponds to a free energy barrier for the association of two reactants, A and B, forming a vibrationally excited species (S$^*$):

$$A + B \xrightarrow{k_f} S^* \qquad (4)$$

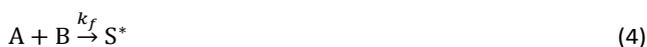

The second bottleneck arises from the competitive formation of the products $P_1$ and $P_2$, occurring near second-order saddle points:

$$S^* \xrightarrow{k_1} P_1 \quad (5)$$

$$S^* \xrightarrow{k_2} P_2$$

The CCUS model enables an approximate of the rate coefficients for the formation of products $P_1$ and $P_2$ as follows:

$$k_j^{\text{CCUS}} = \frac{k_f}{k_f + k_1 + k_2} k_j, \quad j = 1 \text{ or } 2 \quad (6)$$

The association rate coefficient $k_f$ between two reactants with permanent dipole moments can be approximated by:[36]

$$k_f = C\mu^{-1/2}(d_A d_B)^{2/3} T^{-1/6} \quad (7)$$

where $\mu$ is the reduced mass of the two reactants (in amu), $T$ represents the temperature (in K), and $d_A$ and $d_B$ are the dipole moments (in D) of the reactants, calculated via Level2. The constant $C$ has a value of $1.83 \times 10^{-9}$, with units such that $k_f$ is expressed in $cm^3$ $molecule^{-1} s^{-1}$. If one of the reactants does not have a permanent dipole moment, the association rate coefficient is instead given by $C'\mu^{1/2}(d_A Q_B)^{1/2}$, where $Q_B$ is the quadrupole moment (in D Å) and $C' = 3.52 \times 10^{-10}$, with units ensuring that the rate coefficient is expressed in $cm^3$ $molecule^{-1} s^{-1}$.

The rate coefficients $k_1$ and $k_2$ are obtained using canonical variational transition state theory (CVT),[37] with A and B as the reacting species. For processes involving hydrogen transfer, tunneling corrections are included via the small curvature tunneling (SCT)[38] model. The rate coefficients were calculated under the assumption of zero pressure.

When there is no competition between pathways and $S^*$ can undergo only a single unimolecular reaction ($k_1$), the applied model is referred to as the canonical unified statistical (CUS) model, and the rate coefficient is given by $k^{\text{CUS}} = k_f k_1/(k_f + k_1)$.

The vibrationally excited species $S^*$ may also undergo a radiative stabilization process:

$$S^* \xrightarrow{k_r} S + h\nu \quad (8)$$

The rate coefficient for this radiative stabilization was computed using the harmonic approximation:[39]

$$k_r = \sum_{i=1}^{N_m} \sum_{n=0}^{\infty} 1.25 \times 10^{-7} n P_i(n) I_i \nu_i^2 \quad (9)$$

Here, $N_m$ represents the total number of vibrational modes, $P_i(n)$ is the probability of mode $i$ being in level $n$, $I_i$ is the infrared absorption intensity for the transition from level $n = 0$ to $n = 1$ of mode $i$ (in units of km/mol), and $\nu_i$ denotes the frequency of the $i$-th vibrational mode (in $cm^{-1}$). The population distribution can be determined using the following equation:[39]

$$P_i(n, E) = \frac{\rho_{\text{vib}}^{N-1,i}(E - nh\nu_i)}{\rho_{\text{vib}}(E)} \quad (10)$$

In this equation, $\rho_{\text{vib}}^{N-1,i}$ refers to the density of states for the molecule with the $i$-th mode absent.

In the calculations above, all vibrational frequencies were scaled by the recommended factor of 0.975.[40]

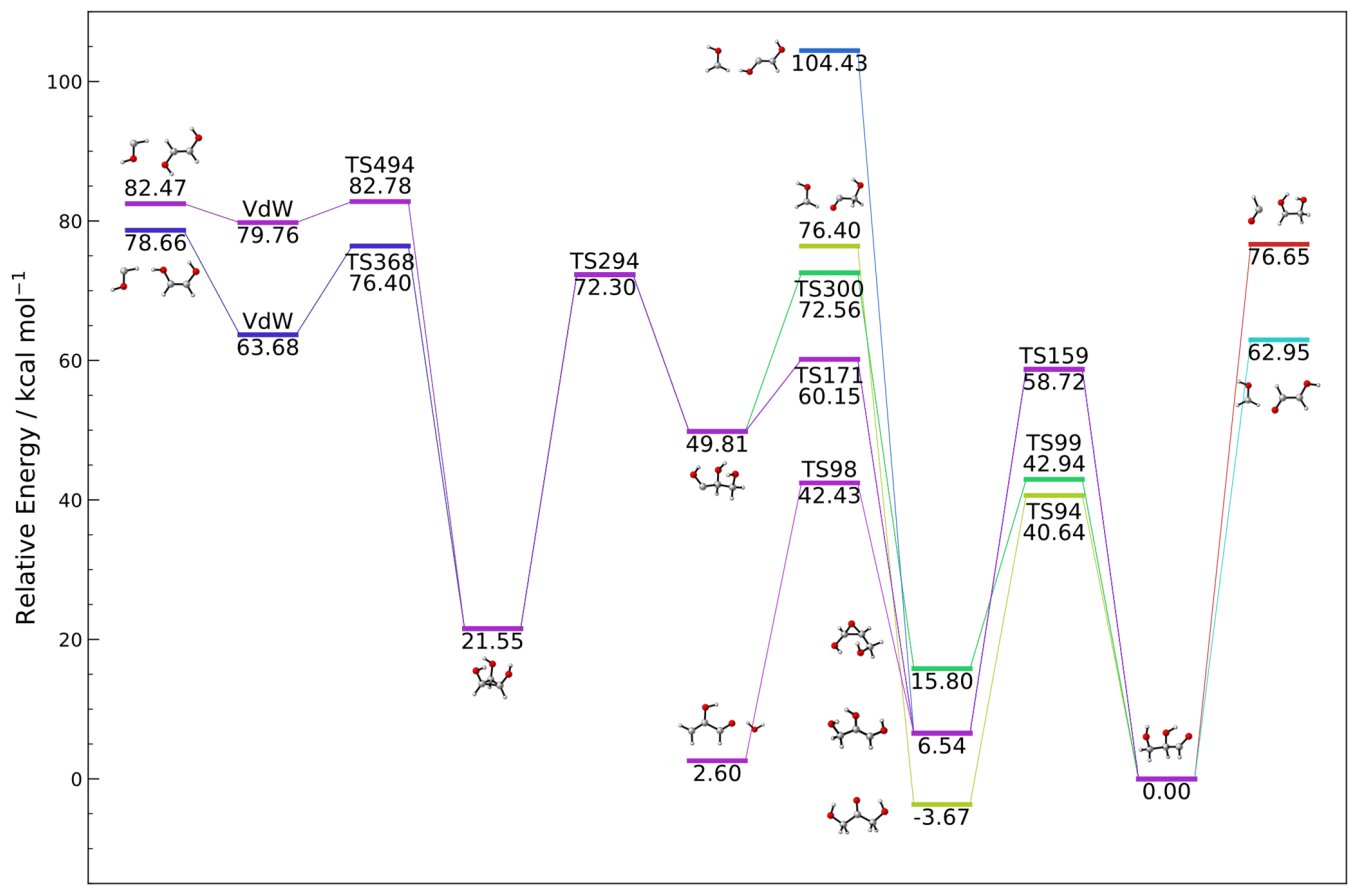

**Fig. 1** DFT-computed energy profile for the most probable formation routes of GCA via radiative association. The energies include zero-point vibrational energy (ZPE), with the zero of energy set for GCA, shown at the bottom right of the figure.

## 3 Results

### 3.1 Analysis of the $C_3H_6O_3$ CRN

The screening of the $C_3H_6O_3$ CRN to find plausible GCA formation pathways consisted of seeking reactions that follow energetically favorable pathways with submerged barriers. This filtering process revealed several viable pathways for the gas-phase formation of glyceraldehyde. Fig. 1 illustrates the energy profile for the most probable and chemically relevant pathways, described from left to right:

1. **Purple and violet-blue pathways (left of Fig. 1).** These routes involve hydroxycarbene (HOCH) reacting with either (*Z*)-ethene-1,2-diol (HOCHCHOH, violet-blue) or (*E*)-ethene-1,2-diol (purple). Among these three species, only (*Z*)-1,2-ethenediol has been detected in the ISM.[10] Notably, the violet-blue route corresponds to the reaction mechanism hypothesized by Rivilla *et al*.[10] These are the only pathways involving a reaction between two singlet species. Both routes lead to GCA in five elementary steps, with the resulting GCA* possessing an excess energy of approximately 80 kcal/mol. The first step in both routes involves the association of reactants to form stable van der Waals (vdW) complexes, which then rearrange to produce cyclopropane-1,2,3-triol (at 21.55 kcal/mol). The rearrangement in the violet-blue route is more favorable than in the purple route.

2. **Dark blue pathway (center of Fig. 1).** This route begins with the barrierless association of two radical species at 104.4 kcal/mol: hydroxymethyl ($CH_2OH$) and HOCCHOH radicals. This pathway converges with the purple one at the (*Z*)-prop-1-ene-1,2,3-triol intermediate (at 6.54 kcal/mol), leading to GCA* in just two steps, but with a significant excess energy of 104 kcal/mol. Additionally, while none of the reactants have been detected in the ISM, protonated formaldehyde ($CH_2OH^+$) has been observed.[41]

   From the (*Z*)-prop-1-ene-1,2,3-triol intermediate, the convergence point of the three aforementioned pathways, the most favorable route proceeds through TS98 to yield 2-hydroxyprop-2-enal ($CH_2COHCHO$) and water, rather than forming GCA.

3. **Light green pathway (center of Fig. 1).** This route shares one reactant with the dark blue pathway: hydroxymethyl radical. The other reactant, the hydroxymethylcarbonyl radical ($HOCH_2CO$), has been detected in the ISM.[42] As in the previous case, this route leads to GCA in two steps.

4. **Red route (right of Fig. 1).** In this pathway, the formyl radical (HCO) undergoes a barrierless radical-radical association with HOCHCH$_2$OH, leading to GCA in a single step. Among the two reagents, only the formyl radical has been detected in the ISM.[43]
5. **Light blue route (right of Fig. 1).** This pathway involves two reactants: the hydroxymethyl radical and the 2-oxoethylideneoxidanium radical (HOCHCHO), the latter of which has been detected in the ISM.[42] Similar to the previous route, GCA forms through the association of the reactants in a single step.

Although some of the reactions described above appear plausible due to their barrierless nature, GCA* forms with a significant excess energy (62.9 kcal/mol in the most favorable case, specifically in the light blue route). In this scenario, GCA* can only dissipate energy through radiative emission of IR light or undergo further unimolecular reactions. An RRKM calculation for the unimolecular decomposition of GCA at 62.9 kca/mol predicts 100% abundance of formaldehyde and ethene-1,2-diol, after surmounting an energy barrier of 36.8 kcal/mol (see Fig. 2). The estimated radiative emission rate is 111 s$^{-1}$, which is orders of magnitude lower than the unimolecular decomposition rate of GCA*, leading to formaldehyde and ethene-1,2-diol, which is approximately 3×10$^6$ s$^{-1}$. Therefore, the following reaction represents the outcome of the radical-radical association described in the light blue route:

$$\mathrm{CH_2OH + HOCHCHO \rightarrow CH_2O + HOCHCHOH} \quad (11)$$

As shown in Fig. 2, there are two TSs with relatively similar energies, one leading to the $Z$ isomer and the other to the $E$ isomer, with the former being 2.4 kcal/mol lower in energy.

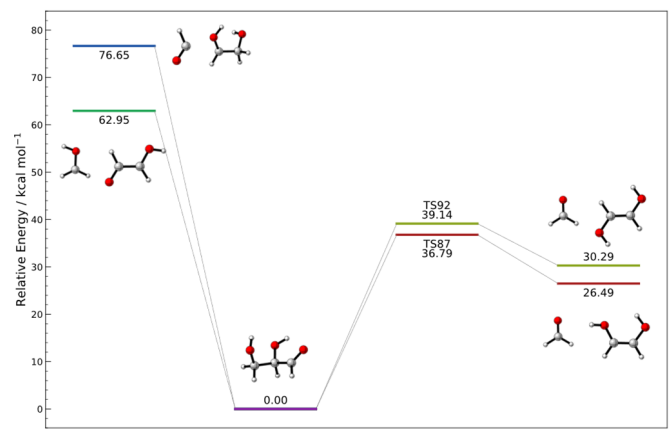

**Fig. 2** DFT-computed energy profile for the reactions CH$_2$OH + HOCHCHO → CH$_2$O + HOCHCHOH and HCO + HOCHCH$_2$OH → CH$_2$O + HOCHCHOH. The energies include zero-point vibrational energy (ZPE), with the zero of energy set at GCA.

The rate coefficients for reaction (11) evaluated using the CCUS model and presented in Table S1 of the SI, indicate that the abundance of the $Z$ isomer is orders of magnitude higher than that of the $E$ isomer. Notably, only the $Z$ isomer has been detected in the ISM.[10]

The rate coefficients for the formation of formaldehyde and (Z)-1,2-ethenediol from reaction (11) in the 10-100 K temperature range vary from $\mathbf{8.7 \times 10^{-10}}$ and $\mathbf{5.9 \times 10^{-10}}$ cm$^3$ molecule$^{-1}$ s$^{-1}$.

Fig. 2, along with our kinetic calculations, shows that the predominant decomposition pathway of GCA leads to the formation of formaldehyde and ethene-1,2-diol. Therefore, the light green, red and light blue pathways depicted in Fig. 1 exclusively yield these two products, with GCA serving as a key intermediate. In contrast, the purple, violet-blue, and dark blue pathways result in the formation of water and 2-hydroxyprop-2-enal.

### 3.2 Analysis of the C$_3$H$_7$O$_3$ CRN

The starting geometry for our analysis with AutoMeKin was a radical formed by GCA with an extra hydrogen atom (GCA-H). The rationale for studying this CRN was to propose that GCA could form according to reaction (2). In this way, the unimolecular decomposition of the system could lead to the formation of GCA, releasing a radical hydrogen, as seen in reaction (2).

As in the CRN described in the previous section, the screening process was similar, *i.e.*, reactions where atomic hydrogen is released and that involved barriers whose energy were below the energy of reactants were sought. Additionally, among the possible reactants, those that could rearrange through an exothermic hydrogen transfer were discarded, as this process

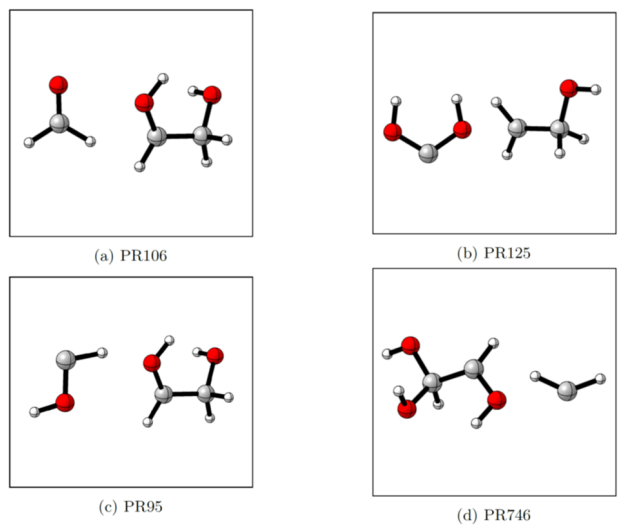

(a) PR106    (b) PR125

(c) PR95    (d) PR746

would potentially be much faster than reaction (2).

**Fig. 3** Potential reactants that undergo a barrierless association leading to GCA-H. Each is labeled using AutoMeKin's notation (PR followed by a number) to facilitate the identification of the structures within the C$_3$H$_7$O$_3$ CRN.

We found that the reactions between the reactant pairs shown in Fig. 3 are particularly relevant.

The reaction between the fragments depicted in Fig. 3a, although leading to the formation of GCA-H without a barrier in a single step, does not yield GCA and H, as their energy is higher than that of the reactants. However, this reaction is of particular interest because no lower-energy pathway exists for these reactants other than its dissociation back into the original fragments. Consequently, the only possible outcomes of a reactive encounter between these two species are either immediate dissociation or stabilization of the complex via spontaneous radiative emission. As mentioned above, formaldehyde is a compound detected in the ISM,[44] whereas the larger fragment, the HOCHCH$_2$OH radical, which also appears in the red route of Fig, 1, has not been observed. However, the reaction between hydroxyacetaldehyde, which has been detected,[45] with atomic hydrogen to form HOCHCH$_2$OH has been reported to be exothermic by 102 kJ/mol, albeit with a significant barrier.[42]

The reactants depicted in Fig. 3b would lead to the formation of GCA + H· in four steps, with the reaction being exothermic by 17.5 kcal/mol. However, neither of these two compounds has been found in the ISM. The two-carbon molecule is the radical derived from ethanol (detected in the ISM)[46] lacking a hydrogen atom. The other reactant in the high-energy isomer of formic acid (HOCOH), which can undergo hydrogen transfer to form the more stable acid.

The reaction between the reactants depicted in Fig. 3c is the most promising and is discussed in more detail below.

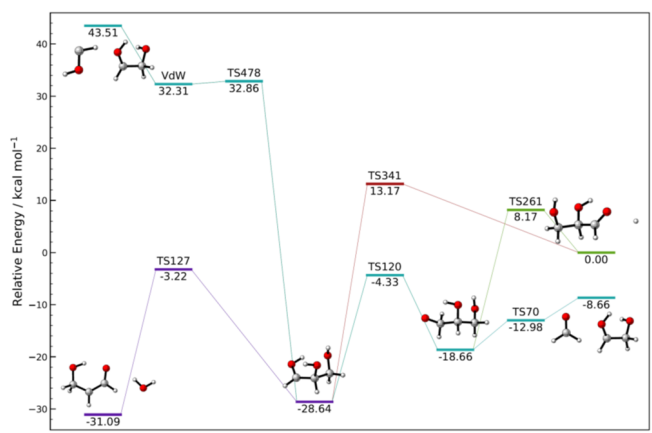

**Fig. 4** DFT-computed energy profile for the reaction HOCH + HOCHCH2OH. The energies include zero-point vibrational energy (ZPE), with the zero of energy set at GCA + H·, shown at the right of the figure.

Finally, the reaction between the fragments shown in Fig. 3d is the most energetically favorable of all, with an energy difference of 60.6 kcal/mol between the reactants and GCA + H·, leading to the formation of GCA in five steps. Additionally, the small fragment, methylene (CH$_2$), has been detected in space.[47] However, the larger fragment is relatively complex, with three oxygen atoms. Nevertheless, the existence of this compound would not be unexpected, as it is structurally similar to ethene-1,2-diol with an additional OH group. Various hypotheses could be proposed regarding its formation.

As shown in Fig. 4, the reaction between hydroxycarbene and HOCHCH$_2$OH (at 43.51 kcal/mol), the reactants of Fig. 3c, is barrierless and highly energetically favorable, leading to the formation of glyceraldehyde and atomic hydrogen after three steps. However, the energy profile of Fig. 4 clearly indicates that the most favorable pathway results in the formation of the fragments displayed in Fig. 3a, formaldehyde and the HOCHCH$_2$OH radical, or to the formation of water and HOCH$_2$CHCHO. The former represents a peculiar and less efficient pathway for the isomerization of hydroxycarbene into formaldehyde.

As mentioned above, hydroxycarbene has not been detected in space. One possible reason for this could be that hydrogen transfer between oxygen and carbon, leading to the more stable formaldehyde conformer, occurs relatively easily. Nevertheless, this compound frequently appears in our CRNs and seems to be a highly reactive species, driving several barrierless and highly exothermic reactions. The numerous reaction pathways that the presence of this compound in the ISM would enable motivated us to study its unimolecular kinetics in greater detail.

The HCOH molecule can adopt two possible conformations: *syn* and *anti*. The *anti* conformation can undergo hydrogen transfer, leading to the formation of formaldehyde as shown in Fig. 5. Using both the CCUS and the CVT/SCT models, the rate coefficient for the formation of formaldehyde from hydroxycarbene in the temperature range 10-50 K is approximately $4 \times 10^{-5}$ s$^{-1}$ (see TableS2 in the SI for detailed values). On the other hand, the rate coefficients for the reaction of hydroxycarbene with HOCHCH$_2$OH radicals are on the order of $7.2-9.4 \times 10^{-10}$ cm$^3$ molecule$^{-1}$ s$^{-1}$. Therefore, the rearrangement of hydroxycarbene depicted in Fig. 5 may potentially compete with its reaction with HOCHCH$_2$OH, depending on the abundance of these radicals.

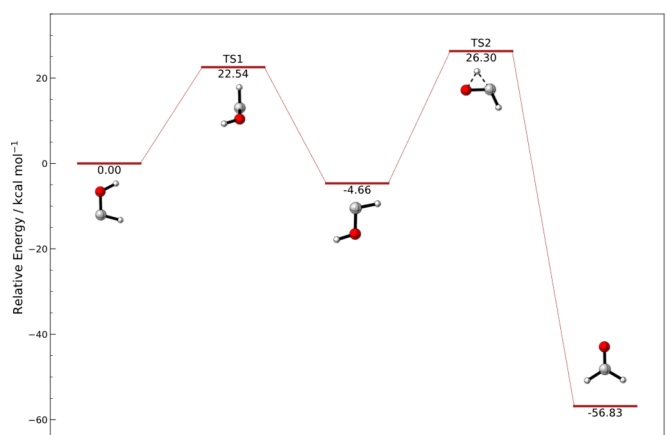

**Fig. 5** DFT-computed energy profile for the reaction HCOH → CH2O rearrangement reaction. The energies include zero-point vibrational energy (ZPE), with the zero of energy set at HCOH.

Finally, we also explored other interesting reactions identified within this CRN, beyond those leading to GCA. In this context, we identified a highly efficient formation route for both hydroxymethyl radical and glycolaldehyde (HOCH$_2$CHO) from hydroxycarbene and the HOCH$_2$CH$_2$O radical:

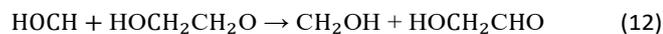

$$\text{HOCH} + \text{HOCH}_2\text{CH}_2\text{O} \rightarrow \text{CH}_2\text{OH} + \text{HOCH}_2\text{CHO} \quad (12)$$

The rate coefficients for this reaction were computed using the CCUS model and found to range from $8.9 \times 10^{-10}$ and $6.1 \times 10^{-10}$ cm$^3$ molecule$^{-1}$ s$^{-1}$ over the 10-100 K temperature interval. See TableS3 in the SI for detailed values.

### 3.3 Analysis of reactions analogous to reaction (2)

The observation that the lowest-energy pathway from GCA-H leads to the formation of formaldehyde and the HOCHCH$_2$OH radical (*vide supra*), prompted us to explore other reaction types. Specifically, we looked at reactions analogous to reaction (2):

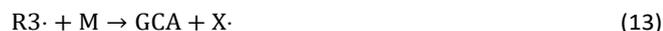

$$\text{R3} \cdot + \text{M} \rightarrow \text{GCA} + \text{X} \cdot \quad (13)$$

where X· is a fragment larger than atomic hydrogen. Eight different reactant combinations were examined, and their reaction energies were computed at the same DFT level employed throughout this study (see Fig. 6). As shown in the figure, the most exothermic and promising reaction occurs between glyoxal (HCOHCO) and the HOCHCH$_2$OH radical, with an exothermicity of 9.2 kcal/mol:

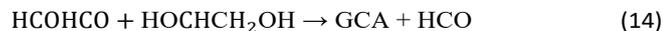

$$\text{HCOHCO} + \text{HOCHCH}_2\text{OH} \rightarrow \text{GCA} + \text{HCO} \quad (14)$$

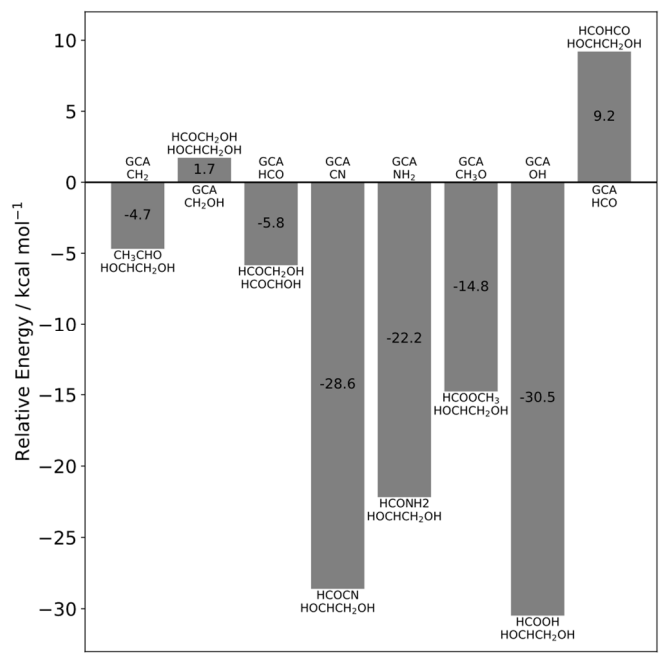

**Fig. 6** DFT-computed reaction energies for various reactions of the type R3· + M → GCA + X (reaction 13, see text). In each case, the zero-energy reference corresponds to the products, with the respective reactant pairs indicated in the figure.

Glyoxal can exist in two structural forms: *E* and *Z*. Although it has not been detected in the ISM yet, its presence in space has been hypothesized based on the dimerization of HCO radicals.[48] One possible reason for its elusive detection is that the lowest-energy isomer (*E*) has a vanishing dipole moment, making it undetectable by rotational emission.[10]

The reaction pathways connecting the reactants and products of reaction (14) were computed at Level2 and are shown in Fig. 7.

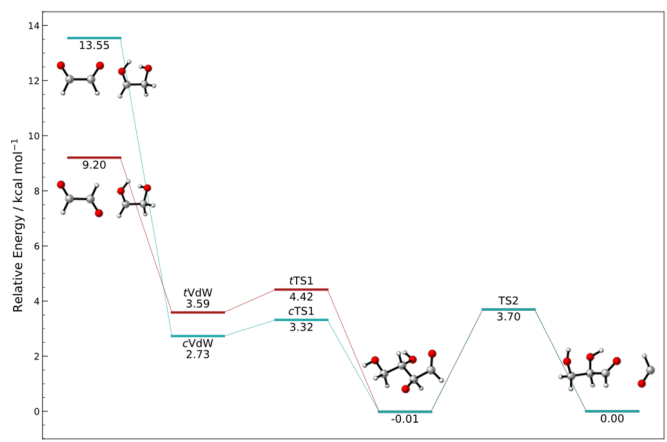

**Fig. 7** DFT-computed energy profile for the reaction HCOHCO + HOCH$_2$CHOH→ GCA + HCO·. The energies include zero-point vibrational energy (ZPE), with the zero of energy set at GCA + HCO·, shown at the right of the figure.

The rate coefficients for the formation of GCA and HCO via reaction (14) were computed in this study. The reaction is limited by the initial association step, and the corresponding rate coefficients are provided in TableS4 of the SI. In the 10-100 K temperature range, the rate coefficients for the *Z* isomer are higher than those for the *E* isomer and range from $5.4 \times 10^{-10}$ to $7.9 \times 10^{-10}$ cm$^3$ molecule$^{-1}$ s$^{-1}$.

### 3.4 Analysis of the C$_2$H$_5$O$_2$ CRN

The radical intermediate HOCHCH$_2$OH appeared in multiple reaction pathways analyzed above. Consequently, we conducted a more detailed investigation of its corresponding CRN. The energy profile highlighting the most relevant pathways is shown in Fig. 8.

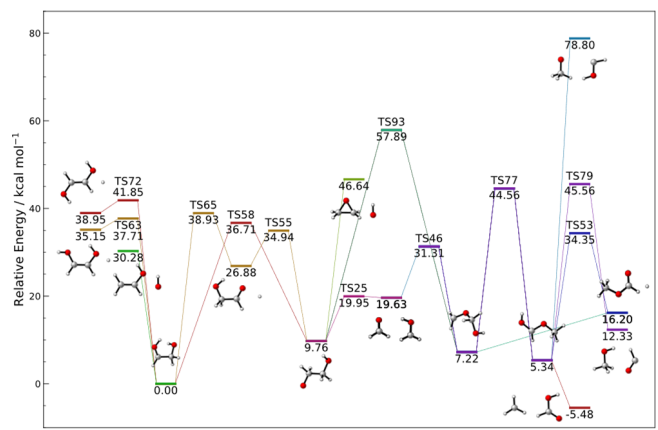

**Fig. 8** DFT-computed energy profile for the most relevant pathways of the C$_2$H$_5$O$_2$ network. The energies include zero-point vibrational energy (ZPE), with the zero of energy set at the HOCHCH$_2$OH radical intermediate.

In the C$_2$H$_5$O$_2$ CRN of Fig. 8, we can distinguish two large regions separated by a high barrier (TS93). On the left side, we observe both the *E* and *Z* isomers of ethene-1,2-diol and atomic hydrogen (at 35.15 and 38.95 kcal/mol). The reaction between any of the two isomers and hydrogen has a barrier, which rules it out as a possible pathway for the formation of other molecules. However, ethene-1,2-diol can be formed from the reaction between oxirane and hydroxyl radical (located in the center of the figure at 46.64 kcal/mol). However, we can assume that this reaction would exclusively lead to the most favorable route, resulting in formaldehyde and the hydroxymethyl radical (at 19.63 kcal/mol).

Another interesting aspect of the profile is the reaction between ethenol (CH$_2$CHOH) and hydroxyl radical (at 30.28 kcal/mol). These two species, both detected in the ISM, react without a barrier to form the HOCHCH$_2$OH radical. Once this radical is formed, the only two possibilities are either decomposition back into the initial reactants or stabilization via spontaneous radiation. This result supports the consideration of the HOCHCH$_2$OH radical as a potential reactant, since, even if it has not been detected, its existence in the ISM would be expected.

On the right side of the energy profile, the other isomer of acetic acid, methyl formate (CH$_3$OCHO), appears along with atomic hydrogen at an energy of 16.20 kcal/mol. Although its formation would be energetically viable from methoxy radical (CH$_3$O) and hydroxycarbene (at 78.80 kcal/mol), it would be expected that formic acid (HCOOH) and methyl radical (at -5.48 kcal/mol) would be the predominant products.

## 4 Conclusions

This study presents a comprehensive computational investigation of the chemical reaction networks associated with the gas-phase chemistry of C$_3$H$_6$O$_3$, C$_3$H$_7$O$_3$, and C$_2$H$_5$O$_2$ species in the context of the interstellar medium. Using the automated reaction discovery tool AutoMeKin in combination with DFT calculations and statistical rate theory, we identified multiple energetically favorable pathways relevant to the formation and decomposition of GCA under astrochemically relevant conditions.

Our analysis of the C$_3$H$_6$O$_3$ CRN reveals several plausible formation routes for GCA, including radical–radical and radical–molecule association reactions. Notably, pathways involving hydroxycarbene, HOCHCH$_2$OH, formyl and hydroxymethyl radicals proceed without barriers and exhibit significant exothermicity. Despite their kinetic feasibility, these reactions typically yield highly energized GCA* molecules, which are prone to rapid unimolecular decomposition before they can be stabilized via radiative cooling. Predicted decomposition products include water, 2-hydroxyprop-2-enal, glycolaldehyde, formaldehyde, and ethene-1,2-diol; all species that have either been observed in the ISM or are considered highly probable.

The C$_3$H$_7$O$_3$ CRN further supports the hypothesis that GCA functions as a transient intermediate. Even within this network, the most accessible pathways favor decomposition over the accumulation of stable GCA. We also explored alternative reactions analogous to GCA formation that involve the departure of larger radicals instead of atomic hydrogen. Among these, the reaction between glyoxal and the HOCHCH$_2$OH radical is particularly notable for its barrierless and exothermic profile. However, the detection of glyoxal in the ISM remains challenging due to its near-zero dipole moment.

Our complementary analysis of the C$_2$H$_5$O$_2$ network underscores the chemical significance of the HOCHCH$_2$OH radical, which emerges repeatedly as an intermediate in multiple reaction schemes. Although this species has not yet been detected in the ISM, its formation is supported by barrierless reactions involving known interstellar molecules such as ethenol and hydroxyl radicals.

Taken together, these results suggest that while the gas-phase formation of GCA is chemically feasible under ISM conditions, it is likely outcompeted by pathways leading to other, more stable complex organic molecules (COMs). GCA may exist only fleetingly in interstellar environments, serving as a reactive intermediate en route to more persistent species. The data generated in this study, including structural information, reaction mechanisms, and rate coefficients, provide essential input for astrochemical models and may inform future observational strategies.

Overall, this work highlights the power of automated CRN exploration in astrochemistry and reinforces the notion that molecular complexity in space emerges not only through stepwise buildup but also through dynamic networks of reactive intermediates, shaped by the unique physical conditions of the interstellar medium.

## Conflicts of interest

There are no conflicts of interest to declare.

## Acknowledgements

This work was partially supported by the Consellería de Cultura, Educación e Ordenación Universitaria (Centro singular de investigación de Galicia acreditación 2019-2022, ED431G 2019/03) and the European Regional Development Fund (ERDF). ALS thanks Xunta de Galicia for financial support through a predoctoral grant. We also thank the Centro de Supercomputación de Galicia (CESGA) for providing access to their computational facilities.

# Supporting Information for:

# Computational Insights into the Chemical Reaction Networks of $C_3H_6O_3$, $C_3H_7O_3$ and $C_2H_5O_2$: Implications for the Interstellar Medium

Anxo Lema-Saavedra,[a] Antonio Fernández-Ramos,[a,b] and Emilio Martínez-Núñez*[b]

[a]Centro Singular de Investigación en Química Biológica y Materiales Moleculares (CIQUS), Universidade de Santiago de Compostela, C/Jenaro de la Fuente s/n, 15782, Santiago de Compostela, Spain

[b]Departamento de Química Física, Facultade de Química, Universidade de Santiago de Compostela, Avda. das Ciencias s/n 15782, Santiago de Compostela, Spain. emilio.nunez@usc.es

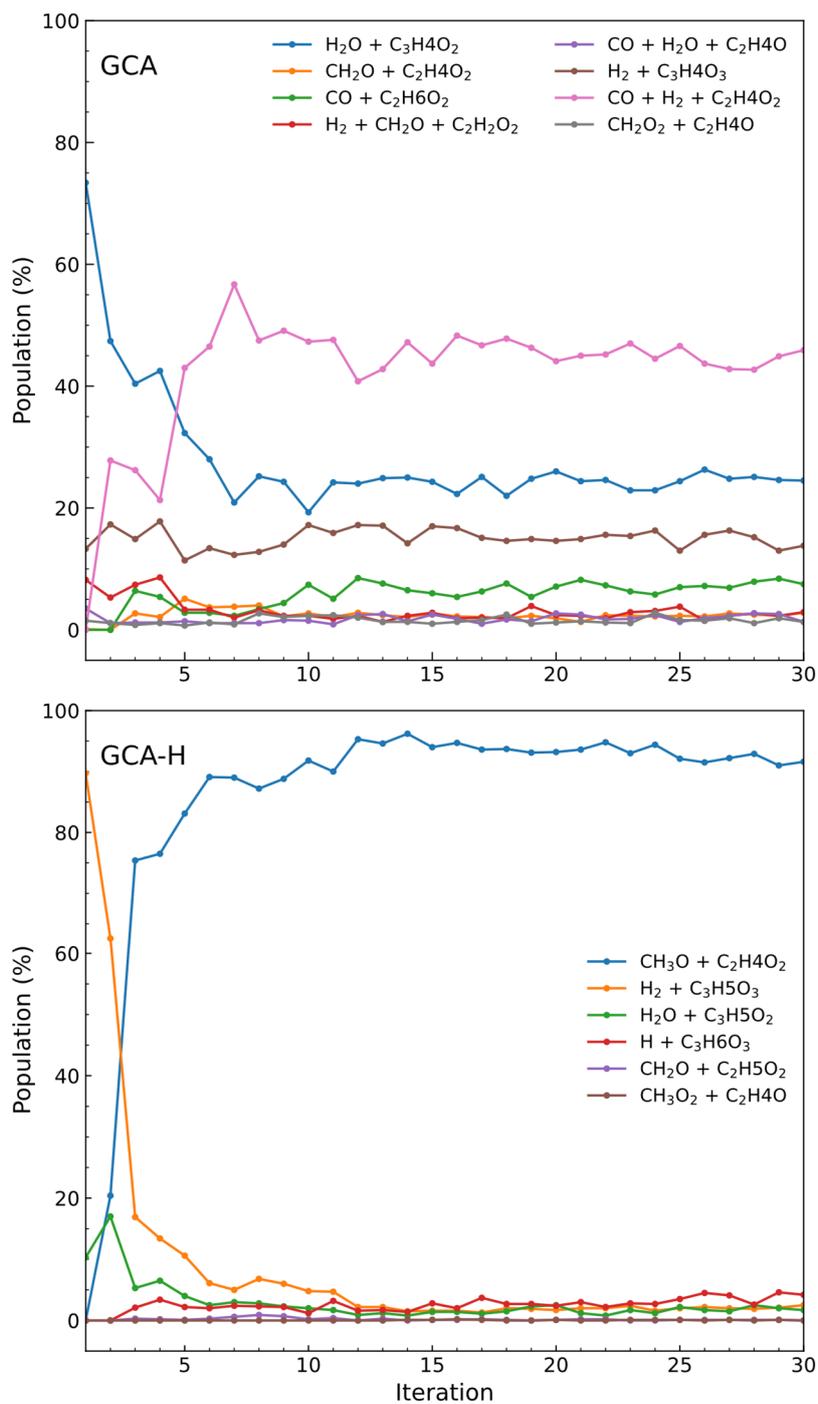

**Fig. S1** Product abundances as a function of number of iterations obtained from the KMC simulations for GCA and GCA-H, using an excitation energy of 250 kcal/mol. The Level1-calculated energies and vibrational frequencies are employed in the calculations of the RRKM rate coefficients, and barrierless reactions are not included.

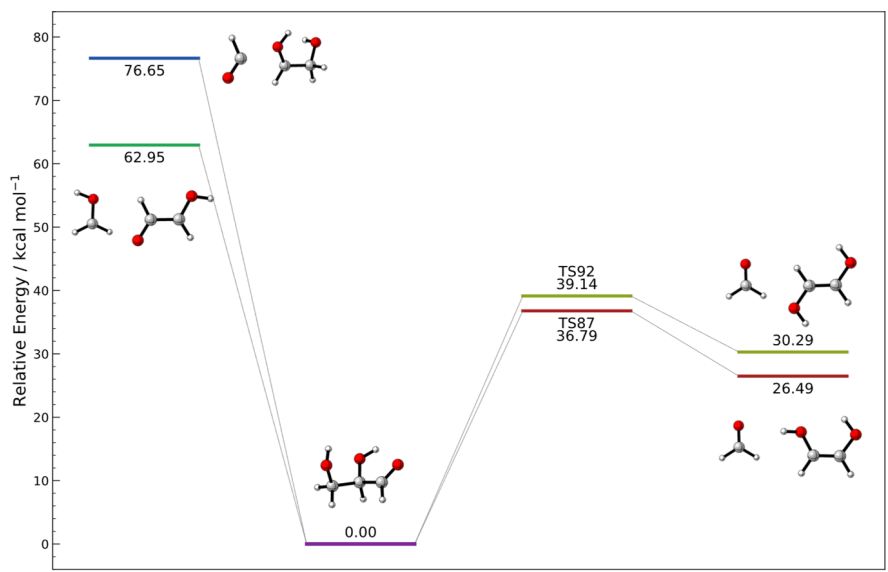

| $T(K)$ | $k_f$ | $k_Z^{TST}$ | $k_E^{TST}$ | $k_Z^{CCUS}$ | $k_E^{CCUS}$ |
|---|---|---|---|---|---|
| 10.00 | 8.753e-10 | 1.919e+558 | 8.906e+506 | 8.753e-10 | 4.062e-61 |
| 15.00 | 8.181e-10 | 2.117e+367 | 1.34e+333 | 8.181e-10 | 5.178e-44 |
| 20.00 | 7.798e-10 | 5.934e+271 | 1.388e+246 | 7.798e-10 | 1.824e-35 |
| 25.00 | 7.513e-10 | 2.501e+214 | 8.086e+193 | 7.513e-10 | 2.429e-30 |
| 30.00 | 7.289e-10 | 1.315e+176 | 1.139e+159 | 7.289e-10 | 6.313e-27 |
| 35.00 | 7.104e-10 | 5.709e+148 | 1.39e+134 | 7.104e-10 | 1.730e-24 |
| 40.00 | 6.947e-10 | 1.659e+128 | 2.781e+115 | 6.947e-10 | 1.165e-22 |
| 45.00 | 6.812e-10 | 1.721e+112 | 7.778e+100 | 6.812e-10 | 3.079e-21 |
| 50.00 | 6.694e-10 | 2.756e+99 | 1.742e+89 | 6.694e-10 | 4.231e-20 |
| 55.00 | 6.588e-10 | 9.219e+88 | 5.057e+79 | 6.588e-10 | 3.614e-19 |
| 60.00 | 6.493e-10 | 1.698e+80 | 5.656e+71 | 6.493e-10 | 2.163e-18 |
| 65.00 | 6.407e-10 | 6.84e+72 | 1.051e+65 | 6.407e-10 | 9.845e-18 |
| 70.00 | 6.329e-10 | 3.115e+66 | 1.779e+59 | 6.329e-10 | 3.615e-17 |
| 75.00 | 6.256e-10 | 9.881e+60 | 1.766e+54 | 6.256e-10 | 1.118e-16 |
| 80.00 | 6.189e-10 | 1.519e+56 | 7.378e+49 | 6.189e-10 | 3.006e-16 |
| 85.00 | 6.127e-10 | 8.567e+51 | 1.008e+46 | 6.127e-10 | 7.209e-16 |
| 90.00 | 6.069e-10 | 1.428e+48 | 3.698e+42 | 6.069e-10 | 1.572e-15 |
| 95.00 | 6.015e-10 | 5.936e+44 | 3.118e+39 | 6.015e-10 | 3.159e-15 |
| 100.00 | 5.963e-10 | 5.362e+41 | 5.333e+36 | 5.963e-10 | 5.931e-15 |

**Table S1** Rate constant values for the formation of formaldehyde and: (*Z*)-1,2-ethenediol or (*E*)-1,2-ethenediol from hydroxymethyl and 2-oxoethylideneoxidanium radicals. The rate constants are given in cm$^3$ molecule$^{-1}$ s$^{-1}$.

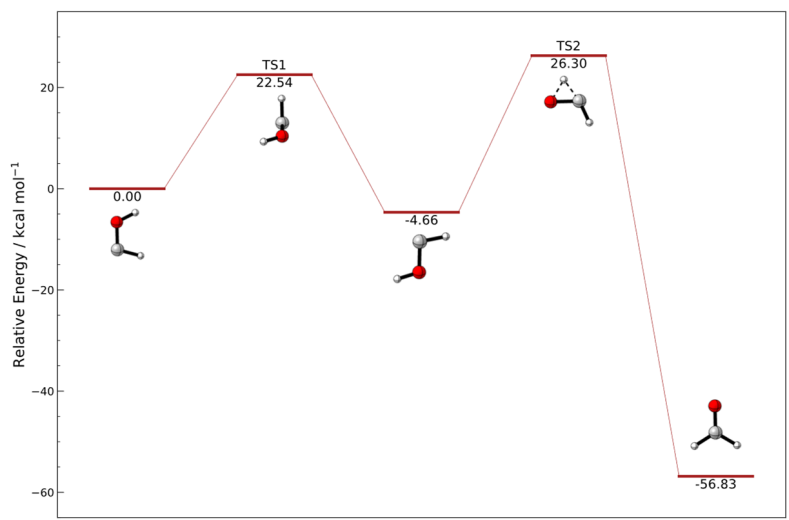

| $T(K)$ | $k^{CCUS}$ (s$^{-1}$) | $k_f$ (cm$^3$ molecule$^{-1}$ s$^{-1}$) |
|---|---|---|
| 10.00 | 3.860e-05 | 9.424e-10 |
| 15.00 | 3.860e-05 | 8.808e-10 |
| 20.00 | 3.860e-05 | 8.396e-10 |
| 25.00 | 3.860e-05 | 8.089e-10 |
| 30.00 | 3.860e-05 | 7.847e-10 |
| 35.00 | 3.860e-05 | 7.648e-10 |
| 40.00 | 3.860e-05 | 7.480e-10 |
| 45.00 | 3.860e-05 | 7.334e-10 |
| 50.00 | 3.860e-05 | 7.207e-10 |

**Table S2** Rate constant values for the transformation of hydroxycarbene into formaldehyde ($k^{CCUS}$), as well as for the association of hydroxycarbene with HOCHCH$_2$OH radicals ($k_f$).

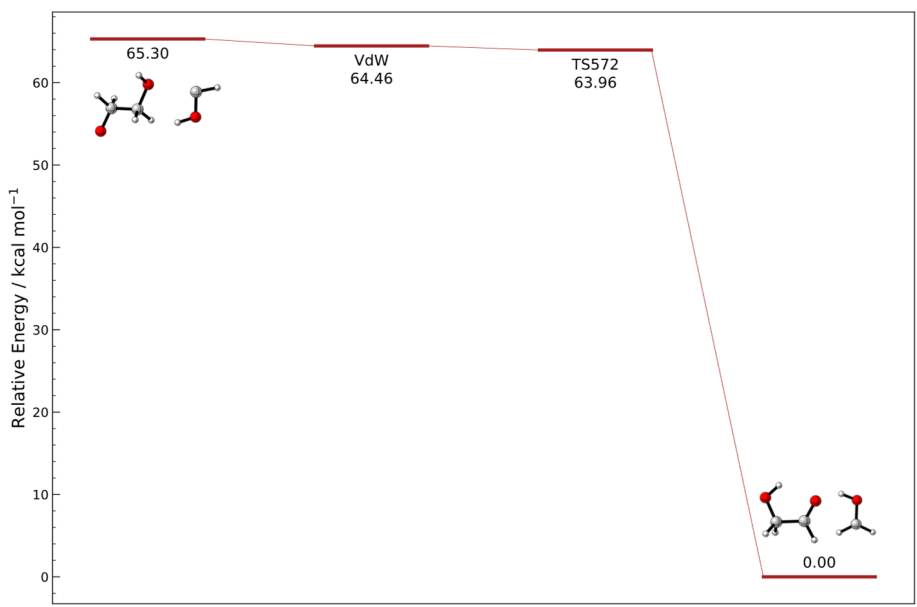

| T(K)   | $k_f$     | $k^{TST}$  | $k^{CUS}$ |
|--------|-----------|------------|-----------|
| 10.00  | 8.988e-10 | 5.519e+16  | 8.988e-10 |
| 15.00  | 8.400e-10 | 3.980e+06  | 8.400e-10 |
| 20.00  | 8.007e-10 | 2.879e+01  | 8.007e-10 |
| 25.00  | 7.715e-10 | 2.180e-02  | 7.715e-10 |
| 30.00  | 7.484e-10 | 1.729e-04  | 7.484e-10 |
| 35.00  | 7.294e-10 | 5.322e-06  | 7.293e-10 |
| 40.00  | 7.133e-10 | 3.856e-07  | 7.120e-10 |
| 45.00  | 6.995e-10 | 4.969e-08  | 6.898e-10 |
| 50.00  | 6.873e-10 | 9.615e-09  | 6.414e-10 |
| 55.00  | 6.765e-10 | 2.506e-09  | 5.327e-10 |
| 60.00  | 6.667e-10 | 8.180e-10  | 3.673e-10 |
| 65.00  | 6.579e-10 | 3.178e-10  | 2.143e-10 |
| 70.00  | 6.498e-10 | 1.416e-10  | 1.163e-10 |
| 75.00  | 6.424e-10 | 7.049e-11  | 6.352e-11 |
| 80.00  | 6.355e-10 | 3.840e-11  | 3.621e-11 |
| 85.00  | 6.291e-10 | 2.253e-11  | 2.175e-11 |
| 90.00  | 6.232e-10 | 1.407e-11  | 1.376e-11 |
| 95.00  | 6.176e-10 | 9.256e-12  | 9.119e-12 |
| 100.00 | 6.123e-10 | 6.368e-12  | 6.302e-12 |

**Table S3** Rate constant values for the formation of the hydroxymethyl radical and glycolaldehyde from HOCH$_2$CH$_2$O and hydroxycarbene. Units are given in cm$^3$ molecule$^{-1}$ s$^{-1}$.

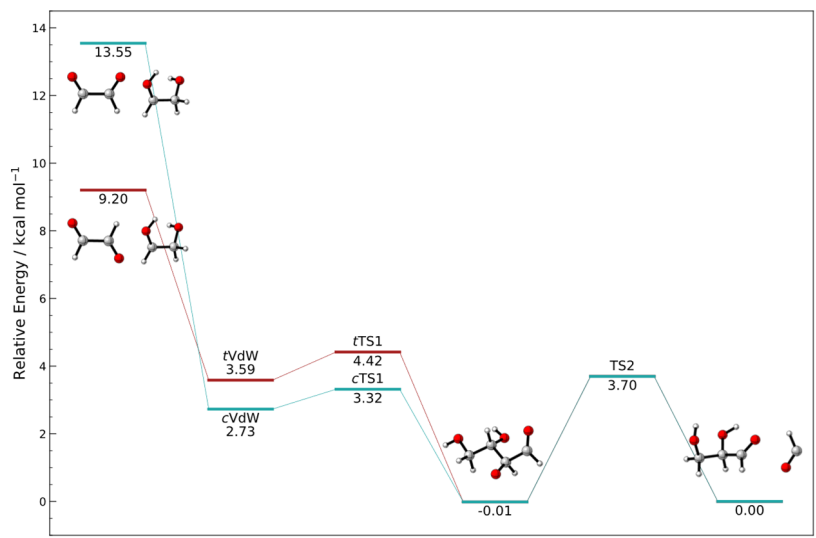

| $T(K)$ | $k_{f,E}$ | $k_E^{CUS}$ | $k_{f,Z}$ | $k_Z^{CUS}$ |
|---|---|---|---|---|
| 10.00 | 2.245e-10 | 2.245e-10 | 7.910e-10 | 7.910e-10 |
| 15.00 | 2.245e-10 | 2.245e-10 | 7.393e-10 | 7.393e-10 |
| 20.00 | 2.245e-10 | 2.245e-10 | 7.047e-10 | 7.047e-10 |
| 25.00 | 2.245e-10 | 2.245e-10 | 6.790e-10 | 6.790e-10 |
| 30.00 | 2.245e-10 | 2.245e-10 | 6.587e-10 | 6.587e-10 |
| 35.00 | 2.245e-10 | 2.245e-10 | 6.420e-10 | 6.420e-10 |
| 40.00 | 2.245e-10 | 2.245e-10 | 6.278e-10 | 6.278e-10 |
| 45.00 | 2.245e-10 | 2.245e-10 | 6.156e-10 | 6.156e-10 |
| 50.00 | 2.245e-10 | 2.245e-10 | 6.049e-10 | 6.049e-10 |
| 55.00 | 2.245e-10 | 2.245e-10 | 5.954e-10 | 5.954e-10 |
| 60.00 | 2.245e-10 | 2.245e-10 | 5.868e-10 | 5.868e-10 |
| 65.00 | 2.245e-10 | 2.245e-10 | 5.790e-10 | 5.790e-10 |
| 70.00 | 2.245e-10 | 2.245e-10 | 5.719e-10 | 5.719e-10 |
| 75.00 | 2.245e-10 | 2.245e-10 | 5.654e-10 | 5.654e-10 |
| 80.00 | 2.245e-10 | 2.245e-10 | 5.593e-10 | 5.593e-10 |
| 85.00 | 2.245e-10 | 2.245e-10 | 5.537e-10 | 5.537e-10 |
| 90.00 | 2.245e-10 | 2.245e-10 | 5.485e-10 | 5.485e-10 |
| 95.00 | 2.245e-10 | 2.245e-10 | 5.436e-10 | 5.436e-10 |
| 100.00 | 2.245e-10 | 2.245e-10 | 5.389e-10 | 5.389e-10 |

**Table S4** Rate constant values for the formation of glyceraldehyde and the HCO radical from the reaction of HOCHCH$_2$OH with glyoxal (both $E$ and $Z$ isomers). Units are given in cm$^3$ molecule$^{-1}$ s$^{-1}$.